\documentclass[aps,pra,twocolumn,superscriptaddress,longbibliography,footinbib]{revtex4-1}  %
\usepackage{graphicx} 
\usepackage{float} 
\usepackage{dcolumn} 
\usepackage{bm,color}
\usepackage{amsmath,amssymb,dsfont,amstext,amsfonts}
\usepackage{extarrows}
\usepackage{xcolor}
\usepackage{wasysym}
\usepackage{mathtools}
\usepackage{bbold}
\usepackage[export]{adjustbox}
\usepackage{mathdots}
\usepackage{siunitx}
\usepackage{comment}
\usepackage[colorlinks=true,linkcolor=blue,urlcolor=blue,citecolor=blue]{hyperref}

\usepackage[normalem]{ulem} 
\usepackage{color}

\begin{document}
\title{Characterization of Spin-Orbit Effects in Superconductors In$_5$Bi$_3$ and In$_5$Sb$_3$}
\author{Yao Wei}
\affiliation{Department of Materials Science and Metallurgy, University of Cambridge, 27 Charles Babbage Road, Cambridge CB3 0FS, United Kingdom}
\affiliation{Theory and Simulation of Condensed Matter
(TSCM), King's College London, Strand, London WC2R 2LS, United Kingdom}
\author{Siyu Chen}
\affiliation{Department of Materials Science and Metallurgy, University of Cambridge, 27 Charles Babbage Road, Cambridge CB3 0FS, United Kingdom}
\affiliation{TCM Group, Cavendish Laboratory, University of Cambridge,
J. J. Thomson Avenue, Cambridge CB3 0HE, United Kingdom}
\author{Bartomeu Monserrat}
\email{bm418@cam.ac.uk}
\affiliation{Department of Materials Science and Metallurgy, University of Cambridge, 27 Charles Babbage Road, Cambridge CB3 0FS, United Kingdom}
\affiliation{TCM Group, Cavendish Laboratory, University of Cambridge,
J. J. Thomson Avenue, Cambridge CB3 0HE, United Kingdom}

\begin{abstract}
We report a first principles computational analysis of two phonon-mediated superconductors, In$_{5}$Bi$_{3}$ and In$_{5}$Sb$_{3}$. We show that spin-orbit coupling leads to splitting of electron bands around the Fermi energy, resulting in a suppression of the electronic density of states in both compounds. In In$_{5}$Bi$_{3}$, the spin-orbit coupling is essential for the dynamical stability of the experimentally observed phase, and the calculated superconducting critical temperature is in close agreement with measurements. In In$_{5}$Sb$_{3}$, the spin-orbit coupling significantly reduces the calculated superconducting critical temperature compared to calculations neglecting relativistic effects. Our work emphasises the subtle interplay between spin-orbit interactions and phonon-mediated superconductivity.
%
\end{abstract}
\maketitle

\section{Introduction}

Over the past century, significant advances in superconductivity research have set the stage for revolutionary applications across diverse sectors. Superconductors are pivotal in magnetic resonance imaging for medical diagnostics\,\citep{lvovsky2013novel}, could facilitate lossless power transmission in electrical grids\,\citep{malozemoff2011electric}, and revolutionise the computational paradigm with quantum computers\,\citep{berggren2004quantum}. Each of these applications underscores the transformative potential of superconductors in shaping modern technology and infrastructure. However, the lack of a superconductor at ambient conditions severely limits their actual impact. 

The development of accurate microscopic theories of superconductivity, together with practical computational frameworks for the evaluation of superconducting critical temperatures, could aid in the search for a superconductor at ambient conditions. Recent progress in modelling techniques and computational capabilities have shed light on various quantum phenomena occurring within superconductor systems\,\citep{van2020roadmap}. These developments are gradually bridging the experiment-theory gap, refining theoretical models that are becoming increasingly predictive. 

In this work, we describe a comparative computational study of two superconductors, In$_{5}$Bi$_{3}$ and In$_{5}$Sb$_{3}$, which have the same crystal structure but different electronic properties. In particular, bismuth exhibits stronger spin-orbit coupling compared to antimony, and we aim to explore the interplay between spin-orbit coupling and superconductivity in these materials. 

The initial discovery of In$_{5}$Bi$_{3}$, dating back to the 1960s\,\citep{wang1969crystal,cruceanu1970electrical,hering1972hall} revealed a critical superconducting transition temperature of 4.2\,K. Subsequent research used a variety of techniques to further characterize superconducting and other properties of In$_5$Bi$_3$\,\citep{fries1982experimental,schreurs1985In5Bi3,bottcher1997investigations,degtyareva2001structural}.
Compared with In$_{5}$Bi$_{3}$, In$_{5}$Sb$_{3}$ has received less attention from the scientific community\,\citep{degtyareva1981crystal,degtyareva1983intermediate,sharma1989sb,shekar1997synthesis}. Specifically, research on the superconductivity properties of In$_{5}$Sb$_{3}$ is limited to two papers from the 1980s, which report a critical superconducting transition temperature of 5.6\,K\,\citep{degtyareva1981crystal,degtyareva1983intermediate}. 

\begin{figure}[ht]
\includegraphics[width=\columnwidth]{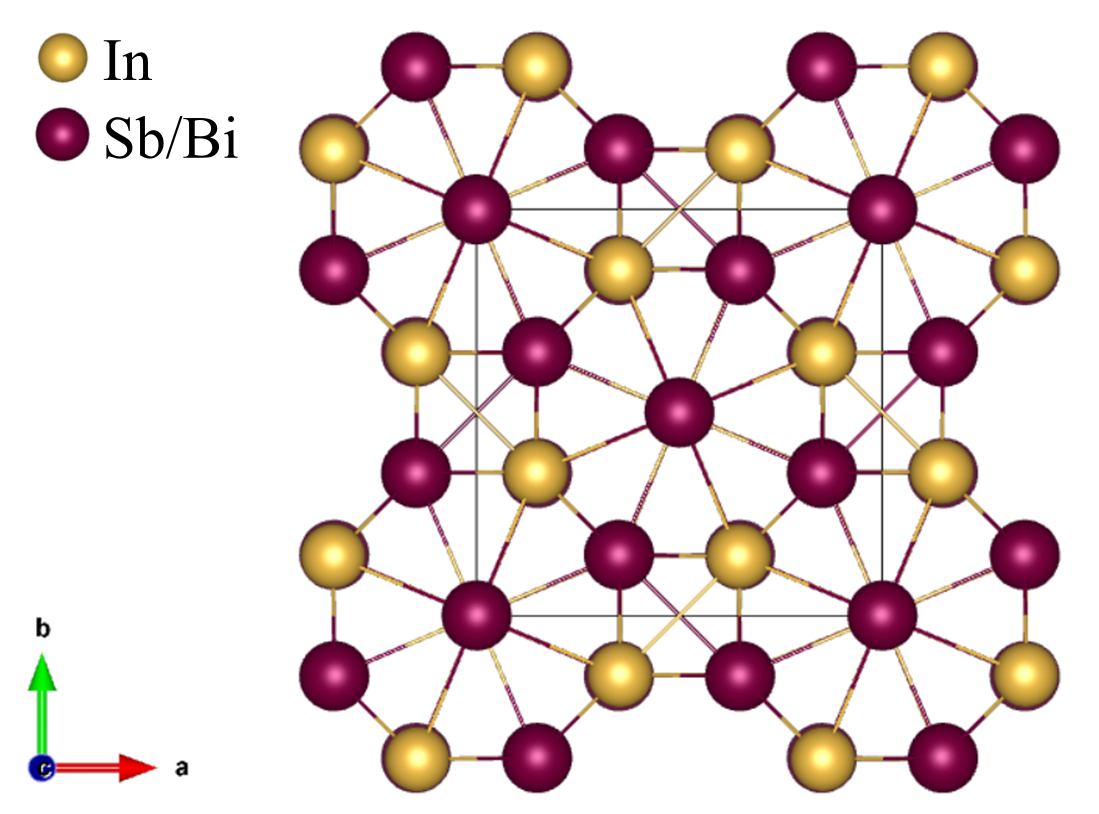}
\caption{Conventional unit cell of In$_{5}$Sb$_{3}$ as reported in entry
640434 of the Inorganic Crystal Structure Database (ICSD)\,\citep{hellenbrandt2004inorganic}.
Similarly, the conventional cell of In$_{5}$Bi$_{3}$ is also categorized
 according to entry 1244 of the ICSD. Both compounds share the same crystallographic space group \textit{I4/mcm} (140) as illustrated in this figure.}
\label{structure}
\end{figure}

In this work, we describe first principles calculations combined with the Migdal-Eliashberg (ME) theory\,\citep{migdal1958interaction,eliashberg1960interactions} to study the superconducting properties of In$_{5}$Bi$_{3}$ and In$_{5}$Sb$_{3}$, with a particular emphasis on the role of spin-orbit coupling (SOC). Regarding In$_{5}$Bi$_{3}$, our calculations confirm the recent proposal that the inclusion of SOC plays a pivotal role in the structural stability of the material, and our predicted superconducting critical temperature is in close agreement with experimental values. For In$_{5}$Sb$_{3}$, our phonon calculations show that the structural stability of the material does not depend on spin-orbit coupling, but the calculated superconducting transition temperature is significantly influenced by spin-orbit coupling effects.

\section{Methods}

\subsection{Structure}

We investigate the physical properties of In$_{5}$Bi$_{3}$ and In$_{5}$Sb$_{3}$ starting from the experimentally determined crystal structures. Both In$_{5}$Bi$_{3}$ and In$_{5}$Sb$_{3}$ have a tetragonal crystal structure with \textit{I4/mcm} space group (No.\,140) and 16 atoms in the primitive cell, as is shown in Fig.\,\ref{structure}. In$_{5}$Bi$_{3}$ has lattice parameters of $a=8.484$\,Å and $c=12.590$\,Å\,\citep{degtyareva1981crystal} and In$_{5}$Sb$_{3}$ has lattice parameters of $a=8.340$\,Å and $c=12.340$\,Å\,\citep{kubiak1977rontgenographische}.

\subsection{Computational methods}

\begin{figure*}[t]
\includegraphics[width=\textwidth]{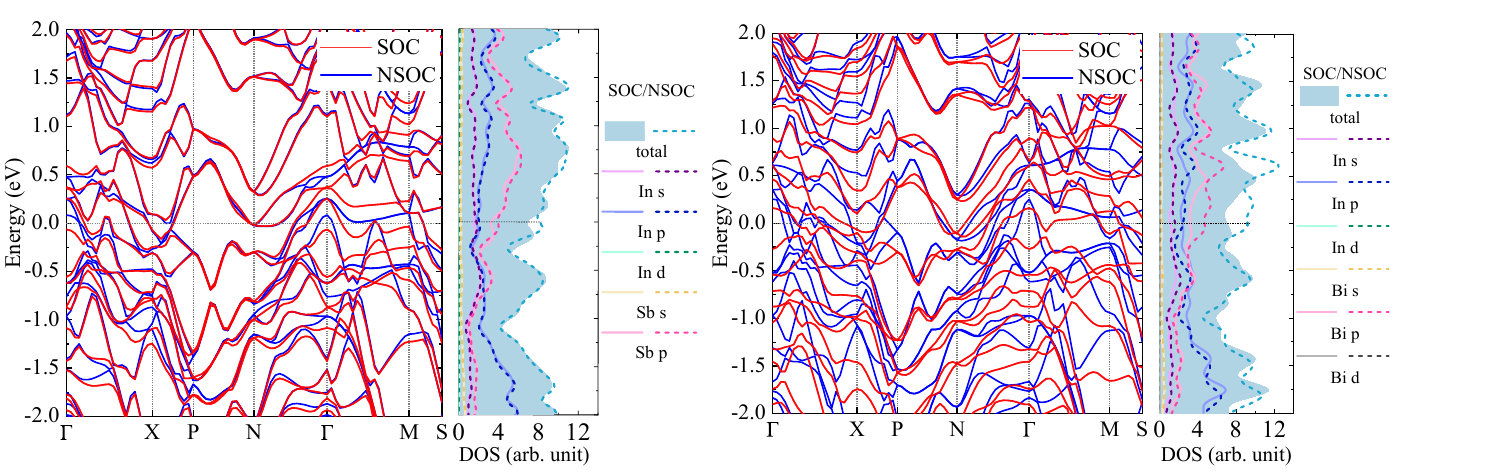}
\caption{Electronic band structure (left) and projected Density of States (DOS,
right) of In$_{5}$Sb$_{3}$ (left panel) and In$_{5}$Bi$_{3}$ (right
panel). The results of non-spin-orbit coupling(NSOC) are shown by blue lines, while
with SOC by red lines. Atom-projected DOS, as well as projections
onto \textit{s}, \textit{p}, \textit{d} orbitals of In are shown in
purple, blue, and green, whereas projections on \textit{s} orbit of
Sb/Bi, \textit{p} orbit of Sb/Bi, \textit{d} orbit of Bi are shown
in yellow, pink and grey, respectively. The zero of the energy is
chosen at the Fermi level.}
\label{band_dos}
\end{figure*}

\begin{figure*}[t]
\includegraphics[width=\textwidth]{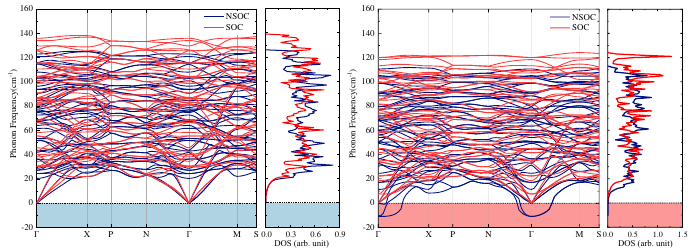}
\caption{Phonon dispersion (left) and phonon DOS (right) of In$_{5}$Sb$_{3}$
(left panel) and In$_{5}$Bi$_{3}$ (right panel). The results with
and without SOC are shown by red and dark blue lines, respectively.}
\label{phonon}
\end{figure*}

\subsubsection{Electronic structure calculations and geometry optimization}

We perform first principles calculations using density functional theory\,\citep{kohn1965self,hohenberg1964inhomogeneous} with \textsc{Quantum
Espresso}\,\citep{giannozzi2009quantum,giannozzi2017advanced}. The generalized gradient approximation
of Perdew-Burke-Ernzerhof (PBE) is used to approximate the exchange-correlation
functional\,\citep{perdew1996generalized}. We use
ultrasoft pseudopotentials\,\cite{vanderbilt1990soft} (pbe-kjpaw-psl) \ taken from the \textsc{Quantum
Espresso} pseudopotential library generated by Dal Corso\,\citep{dal2014pseudopotentials}.  The valence electrons are from the following atomic shells: indium $4d^{10}5s^{2}5p^{1}$, bismuth $5d^{10}6s^{2}6p^{3}$, and antimony $5s^{2}5p^{3}$.
After convergence tests, we choose a kinetic energy cutoff of $60$\,Ry,
 and a Monkhorst-Pack $\mathbf{k}$-point
grid of size $6\times6\times6$ for both In$_{5}$Bi$_{3}$ and In$_{5}$Sb$_{3}$. Moreover, when the total energy change between two consecutive self-consistent iterations is less than 2\,$\times$\,$10^{-9}$\,Ry, the calculation is considered as converged. We report calculations
both with and without the spin-orbit coupling (SOC)\,\citep{winkler2003spin}. The ionic relaxation is considered complete if atomic forces are less than $10^{-10}$\,Ry/bohr, and we keep the cell parameters fixed to their experimental values. 

\subsubsection{Phonons and electron-phonon coupling}

We calculate phonons using density functional perturbation theory (DFPT) as implemented in {\sc Quantum Espresso}\,\citep{baroni2001phonons,giannozzi2009quantum}.
We use a coarse 4\,$\times$\,4\,$\times$\,4 phonon $\mathbf{q}$-point grid to evaluate the force constants, and then use Fourier interpolation of the dynamical matrices onto a finer grid to evaluate the phonon dispersion. 

For the electron-phonon coupling calculations, we employ a $6\times6\times6$ electron $\mathbf{k}$-point grid and a $4\times4\times4$ phonon $\mathbf{q}$-point grid to compute the electron-phonon matrix elements in {\sc Quantum Espresso}\,\citep{baroni2001phonons,giannozzi2009quantum}. Subsequently, we used the \textsc{wannier90} package \citep{mostofi2008wannier90,mostofi2014updated} and the \textsc{epw} package \cite{giustino2007electron,margine2013anisotropic,ponce2016epw} to interpolate the electron-phonon coupling (EPC) constants onto coarse $8\times8\times8$ $\mathbf{k}$-point and $4\times4\times4$ $\mathbf{q}$-point meshes, as well as onto finer $24\times24\times24$ $\mathbf{k}$-point and $8\times8\times8$ $\mathbf{q}$-point meshes.
The initial guess for the projections on localized orbitals using \textsc{wannier90} is detailed in the Supporting Information.

\subsubsection{Superconducting transition temperature}

We evaluate the superconducting critical temperature  $T_{\mathrm{c}}$ using the Allen-Dynes formula\,\citep{allen1975transition,dynes1972mcmillan}:
\begin{equation}
T_{\mathrm{c}}=\frac{\omega_{\log}}{1.2}\exp\left(-\frac{1.04(1+\lambda)}{\lambda-\mu^{*}(1+0.62\lambda)}\right)\,,
\end{equation}
where $\omega_{\log}=\exp\left[\frac{2}{\lambda}\int\frac{d\omega}{\omega}\alpha^{2}F(\omega)\log\omega\right]$ is a logarithmic average of the phonon frequencies\,\citep{richardson1997high,lee1995first}, $\mu^{*}$ is the screened Coulomb potential that we treat as an empirical parameter\,\citep{margine2013anisotropic}, $\lambda=2\mathop{\int}\nolimits _{0}^{\infty}\frac{{\alpha^{2}F\left(\omega\right)}}{\omega}d\omega$ is the electron-phonon coupling constant, and $\alpha^{2}F\left(\omega\right)$ is the Eliashberg spectral function. The spectral function is given by\,\citep{allen1972neutron,allen1975transition}:
\begin{equation}
\alpha^{2}F(\omega)=\frac{1}{2\pi N\left(\epsilon_{F}\right)}\sum_{\mathbf{q}v}\delta\left(\omega-\omega_{\mathbf{q}v}\right)\frac{\gamma_{\mathbf{q}\nu}}{\hbar\omega_{\mathbf{q}\nu}}\,,
\end{equation}
where $N\left(\epsilon_{F}\right)$ is the electronic density of states at the Fermi level, $\omega_{\mathbf{q}v}$ is the frequency of the phonon with wave vector $\mathbf{q}$ and branch label $\nu$, and $\gamma_{\mathbf{q}\nu}$ is the associated phonon linewidth arising from electron-phonon coupling. In turn, the linewidth is given by\,\citep{allen1972neutron,allen1975transition}:

\begin{equation}
\gamma_{\mathbf{q}\nu} = \frac{2\pi \omega_{\mathbf{q}\nu}}{\Omega_{\mathrm{BZ}}} \sum_{ij} \int d^{3}k \left| g_{\mathbf{k},\mathbf{q}\nu}^{ij} \right|^{2} \delta\left( \epsilon_{\mathbf{k},i} - \epsilon_{F} \right) \delta\left( \epsilon_{\mathbf{k}+\mathbf{q},j} - \epsilon_{F} \right),
\end{equation}
where $\epsilon_{\mathbf{k},i}$ is the electronic energy for band
$i$ and wave vector $\mathbf{k}$, $\Omega_{\mathrm{BZ}}$ is the volume of the Brillouin zone, and $g_{\mathbf{k},\mathbf{q}\nu}^{ij}$ is the electron-phonon interaction matrix element. Finally, the electron-phonon matrix element is given by:
\begin{equation}
g_{\mathbf{k},\mathbf{q}\nu}^{ij}=\left(\frac{\hbar}{2M\omega_{\mathbf{q}\nu}}\right)^{1/2}\left\langle \psi_{i\mathbf{k}+\mathbf{q}}\left|\partial_{\mathbf{q}\nu}V\right|\psi_{j\mathbf{k}}\right\rangle \,,
\end{equation}
where $M$ is the mass of the atoms, $\psi_{i\mathbf{k}+\mathbf{q}}$
and $\psi_{j\mathbf{k}}$ represent the Bloch wavefunctions of the
electronic states involved in the coupling, and $\partial_{\mathbf{q}\nu}V$
denotes the derivative of the crystal potential with respect to the
$\mathbf{q}\nu$ phonon displacement.

We also estimated $T_{\mathrm{c}}$ by solving the Eliashberg equations\,\cite{allen1983theory,margine2013anisotropic,marsiglio1988iterative}:
\begin{equation}
Z\left(i \omega_n\right)=1+\frac{\pi T}{\omega_n} \sum_{n^{\prime}} \frac{\omega_{n^{\prime}}}{\sqrt{R\left(i \omega_{n^{\prime}}\right)}} \lambda\left(n-n^{\prime}\right),    
\end{equation}
\begin{equation}
    Z\left(i \omega_n\right) \Delta\left(i \omega_n\right)=\pi T \sum_{n^{\prime}} \frac{\Delta\left(i \omega_{n^{\prime}}\right)}{\sqrt{R\left(i \omega_{n^{\prime}}\right)}}\left[\lambda\left(n-n^{\prime}\right)-\mu^{*}\right],
\end{equation}
where \( R\left(i \omega_{n^{\prime}}\right) \) and \( \lambda\left(n-n^{\prime}\right) \) are given by:
\begin{equation}
    R\left(i \omega_n\right)=\omega_n^2+\Delta^2\left(i \omega_n\right),
\end{equation}
\begin{equation}
\lambda\left(n-n^{\prime}\right)=\int_0^{\infty} d \omega \frac{2 \omega \alpha^2 F(\omega)}{\left(\omega_n-\omega_{n^{\prime}}\right)^2+\omega^2},
\end{equation}
and where $\omega_n$ stands for the Matsubara frequency, and $Z$ and $\Delta$ represent the renormalization function and the superconducting gap, respectively.

\section{Results and Discussion}

\begin{figure*}[t]
\includegraphics[width=\textwidth]{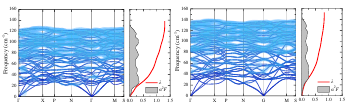}
\caption{Phonon linewidth, the width of the lines is proportional to their intrinsic linewidth (left) and Eliashberg function $\alpha^{2}F$ and
accumulated EPC coefficient $\lambda$ (right) for In$_{5}$Sb$_{3}$
(left panel) without SOC and (right panel) with SOC.}
\label{insb_linewidth}
\end{figure*}

\begin{figure}[t]
\includegraphics[width=\columnwidth]{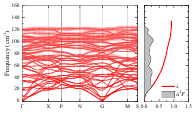}
\caption{Phonon linewidth, the width of the lines is proportional to their intrinsic linewidth (left) and Eliashberg function $\alpha^{2}F$ and
accumulated EPC coefficient $\lambda$ (right) for In$_{5}$Bi$_{3}$
with SOC.}
\label{inbi_linewidth}
\end{figure}

\subsection{Electronic properties}

We present the electronic band structure and the atom-projected density
of states (PDOS) with and without SOC for both In$_{5}$Sb$_{3}$
and In$_{5}$Bi$_{3}$ in Fig.\,\ref{band_dos}. Both compounds exhibit
metallic band structures with similar band dispersions.

The effects of SOC on the band structure are more pronounced in In$_{5}$Bi$_{3}$, which is expected from the larger spin-orbit coupling effect of the
heavier bismuth compared to that of antimony. Indeed, the SOC electronic
structure of In$_{5}$Bi$_{3}$ exhibits an important reduction in
the density of states near the Fermi level compared to the non-SOC
counterpart, a feature that has been shown to be key in determining
the structural stability of this compound\,\citep{chen2019chemical}. Our PDOS results clarify that the main disparity between calculations of In$_{5}$Bi$_{3}$ with and without SOC originates from the distinct p-orbital characteristics of bismuth.

\subsection{Phonon properties}

The calculated phonon spectra and phonon DOS for In$_{5}$Sb$_{3}$
and In$_{5}$Bi$_{3}$, with and without SOC, are shown in Fig.\,\ref{phonon}.
In$_{5}$Sb$_{3}$ is dynamically stable irrespective of whether SOC
is included or not. The inclusion of SOC causes an upward shift in
phonon frequencies across the spectrum, with the highest frequency
obtained with SOC surpassing the non-SOC counterpart by 15\,cm$^{-1}$.

In$_{5}$Bi$_{3}$ exhibits an imaginary frequency phonon mode centered
around the $\Gamma$ point, when the calculation is performed without
the SOC, but the phonon frequencies become real with the inclusion
of the SOC. This finding is consistent with the results reported in
Ref.\,\citep{chen2019chemical}, and indicates that the dynamical
stability of In$_{5}$Bi$_{3}$ requires the inclusion of the spin-orbit
coupling. We note that the highest phonon frequency in the present
calculations is higher in the SOC-inclusive case compared to those
conducted without SOC, which is an opposite trend to that reported
previously\,\citep{chen2019chemical}. This discrepancy likely arises
from differences in the first principles implementations used, including
different pseudopotentials and different treatment of the spin-orbit
interaction.

A comparison of the phonon dispersions of In$_{5}$Sb$_{3}$ and In$_{5}$Bi$_{3}$
shows that the phonon frequencies for In$_{5}$Sb$_{3}$ exhibit higher
values than those of In$_{5}$Bi$_{3}$, as expected from the smaller
mass of antimony compared to bismuth.

\subsection{Electron-phonon coupling}

\begin{figure*}[t]
\includegraphics[width=\textwidth]{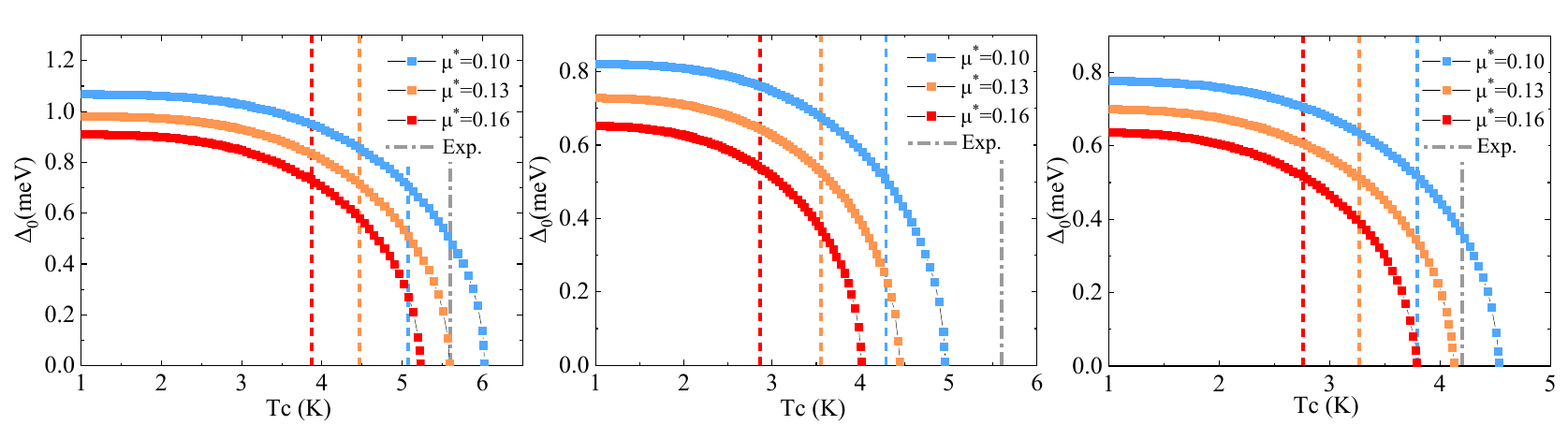}
\caption{The isotropic gap of stable In$_{5}$Sb$_{3}$ and In$_{5}$Bi$_{3}$
as a function of temperature: (a) In$_{5}$Sb$_{3}$ without SOC,
(b) In$_{5}$Sb$_{3}$ with SOC and (c) In$_{5}$Bi$_{3}$ with SOC.
The dashed vertical
lines show the  $T_{\mathrm{c}}$ via the Mcmillan equation with $\mu^{*}$
taken as 0.16, 0.13 and 0.10, respectively.}
\label{tc}
\end{figure*}

\subsubsection{Linewidth}

Figures\,\,\ref{insb_linewidth} and \ref{inbi_linewidth} show
the calculated electron-phonon coupling linewidth, Eliashberg function $a^{2}F$ and
cumulative EPC coefficient $\lambda$ for In$_{5}$Sb$_{3}$
and In$_{5}$Bi$_{3}$. 

In Fig.\,\ref{insb_linewidth}(a), the phonon
dispersion and phonon linewidth of In$_{5}$Sb$_{3}$ without
the SOC are presented. It can be observed that in regions where the
phonon frequencies are higher, the linewidth tends to be larger as
well. Figure\,\ref{insb_linewidth}(b) gives the phonon dispersion and
phonon linewidth for In$_{5}$Sb$_{3}$ with SOC included. As previously
mentioned, the phonon dispersion exhibits higher frequencies in the
presence of SOC compared to the one without SOC. Despite the changes
in the dispersion, the phonon linewidth maintains a strikingly similar
trend when SOC is introduced. This observation indicates that the
effect of SOC on the phonon linewidth is minimal, with only a 1.05\,$\times$\,10$^{-2}$
meV difference in the maximum phonon linewidth observed between the two cases with
and without SOC. This finding might
be significant as it suggests that, at least in the context of In$_{5}$Sb$_{3}$,
the inclusion of the spin-orbit coupling has a more pronounced effect
on phonon frequencies rather than on phonon linewidth.

In the case of In$_{5}$Bi$_{3}$ the phonon spectrum
without SOC contains imaginary frequencies and hence only results
with the SOC included have physical significance. Figure~\ref{inbi_linewidth}
shows the phonon dispersion and phonon linewidth characteristics for
In$_{5}$Bi$_{3}$ with SOC. It reveals
a distinct behaviour compared to the case of the In$_{5}$Sb$_{3}$
system: the phonon linewidth of In$_{5}$Bi$_{3}$
is lower than the corresponding value in In$_{5}$Sb$_{3}$ under
the same SOC conditions. The reason lies in the greater atomic mass
of bismuth in comparison to antimony. This results in a diminished elastic force between atoms,
subsequently leading to a weaker electron-phonon interaction and a reduced
value of the phonon linewidth for In$_{5}$Bi$_{3}$.

\subsubsection{Superconducting transition temperature}

In Fig.\,\ref{tc}, the isotropic gaps of In$_{5}$Sb$_{3}$ and In$_{5}$Bi$_{3}$ are shown as a function of temperature. These results are obtained by solving the Eliashberg equations for each temperature while considering a few typical values of the effective screened Coulomb repulsion constant $\mu^{*}$ in the range of 0.1-0.16\,\citep{lee1995first}. Additionally, the McMillan equation is used to estimate the superconducting critical temperature for comparison.

Figure\,\,\ref{tc}(a)
displays the  $T_{\mathrm{c}}$ prediction for In$_{5}$Sb$_{3}$ without
SOC. At 0\,K, the gap is observed with isotropic characteristics, and the energy of this gap spreads ranging from 0.9 to 1.1\,meV when $\mu^{*}$
takes values from 0.10 to 0.16, the gap vanishes at $5.2-6.0$\,K,
which is identified as the  $T_{\mathrm{c}}$ from our calculation.
 Furthermore, the  $T_{\mathrm{c}}$
was also estimated using the Allen-Dynes modified McMillan equation,
with the same $\mu^{*}$ values from 0.10 to 0.16. The calculated
 $T_{\mathrm{c}}$ values are from $3.87$\,K to $5.07$\,K; these
predictions are lower than the  $T_{\mathrm{c}}$ values obtained from
the Eliashberg equations.

In Fig.\,\ref{tc}(b), we investigate the impact of SOC on the superconducting
transition temperature of In$_{5}$Sb$_{3}$. The inclusion of SOC leads to a reduction in the predicted  $T_{\mathrm{c}}$
when compared to the scenario without SOC shown in panel (a). At 0\,K,
the energy range of the superconducting gap spans from 0.65 to 0.81\,meV. Under the influence of SOC, the
critical temperature, which signifies the onset of superconductivity,
is estimated to be approximately 4.94\,K, 4.48\,K and 3.98\,K when
$\mu^{*}$ takes values of 0.10, 0.13 and 0.16, respectively, as calculated
using the Eliashberg equations. Furthermore, alternative calculations
utilizing the Allen-Dynes modified McMillan equation produce  $T_{\mathrm{c}}$
predictions of 4.29\,K, 3.56\,K and 2.86\,K when $\mu^{*}$ is
equal to 0.10, 0.13 and 0.16, respectively.

Comparing the results obtained with and without SOC for In$_{5}$Sb$_{3}$, we observed a higher superconducting transition temperature in the non-SOC case. This trend can be attributed to a reduction in the electronic density of states at the Fermi level, particularly in the antimony \textit{p} orbitals, upon including SOC. This reduction may be a general phenomenon, as SOC typically lifts band degeneracies, which can reduce the electronic density of states near the Fermi level.

As illustrated in Fig.\,\ref{phonon}, In$_{5}$Bi$_{3}$ exhibits imaginary frequencies in the absence of SOC, so it is not meaningful
to discuss the  $T_{\mathrm{c}}$ under these conditions. Instead, we can perform a calculation in the presence of SOC, and in Fig.\,\ref{tc}(c) the dependence of the gap on temperature and
the calculated values of  $T_{\mathrm{c}}$ for In$_{5}$Bi$_{3}$
for a range of $\mu^{*}$ values are shown. Predictions of  $T_{\mathrm{c}}$
closely resemble those of In$_{5}$Sb$_{3}$ with the inclusion of
SOC when both the modified Allen-Dynes McMillan equation and the Eliashberg
equation are used. Using the Eliashberg equation,  $T_{\mathrm{c}}$
is estimated to be approximately $4.51$\,K, $4.14$\,K
and $3.78$\,K 
when $\mu^{*}$ is set to 0.10, 0.13 and 0.16, respectively. Alternative
calculations utilizing the Allen-Dynes modified McMillan equation
produce  $T_{\mathrm{c}}$ predictions of $3.79$\,K, $3.27$\,K
and $2.76$\,K when $\mu^{*}$ is set to 0.10, 0.13 and 0.16, respectively,
which somewhat underestimates the value from Eliashberg
equation. 

Notably, the experimental data reported in a previous study\,\citep{chen2019chemical}
recorded a $T_{\mathrm{c}}$ of approximately $4.2$\,K for In$_{5}$Bi$_{3}$ which falls within the predicted range and is closest to the value we obtain with $\mu^{*}=0.13$. This agreement between theoretical
calculations and experiments can be considered excellent. 
In contrast, earlier experimental studies \citep{degtyareva1981crystal} have reported a superconducting transition temperature for In$_{5}$Sb$_{3}$ in the range of approximately 5.0-5.6 K. Our predictions under spin-orbit SOC conditions fall short of this experimentally observed range. It is also noteworthy that this measurement was documented only once, nearly forty years ago.

\section{Conclusions}

In summary, our investigation into the electronic and phonon properties
of In$_{5}$Sb$_{3}$ and In$_{5}$Bi$_{3}$ has provided valuable
insights into their superconducting behaviour and the impact
of the spin-orbit coupling on their electronic and phonon structures
and on their superconducting properties. 
Our analysis demonstrates that spin-orbit coupling can exert a significant influence on these properties,
leading to shifts in phonon frequencies and superconducting transition
temperatures. Specifically, we find that the inclusion of spin-orbit coupling reduces the electronic density of states near the Fermi energy, which drives the structural stabilisation of In$_{5}$Bi$_{3}$ and suppresses the superconducting critical temperature in In$_5$Sb$_3$. These results
highlight the importance of accounting for spin-orbit coupling in predictive models,
as it can have a substantial impact on the physical properties of materials containing heavy elements.

The theoretical predictions for the critical temperature of In$_{5}$Bi$_{3}$
demonstrate a close alignment with experimental results,
which underscores the validity of
our theoretical framework for explaining the superconducting characteristics of such materials. The predicted superconducting temperature of In$_{5}$Sb$_{3}$ somewhat underestimates the experimentally reported value, but the latter has only been reported once decades ago, so it may be interesting to revisit the superconducting properties of In$_{5}$Sb$_{3}$ experimentally. 


\section*{Acknowledgments}
Y.W. gratefully acknowledges funding from the China Scholarship Council. S.C. acknowledges financial support from the Cambridge Trust and from the Winton Programme for the Physics of Sustainability. B.M. acknowledges support from a UKRI Future Leaders Fellowship (MR/V023926/1), from the Gianna Angelopoulos Programme for Science, Technology, and Innovation, and from the Winton Programme for the Physics of Sustainability. Computational resources were provided through our membership in the UK's HEC Materials Chemistry Consortium, funded by EPSRC (EP/R029431 and EP/X035859). This work utilized ARCHER2, the UK National Supercomputing Service (http://www.archer2.ac.uk), as well as resources from the UK Materials and Molecular Modelling Hub (MMM Hub), which is partially supported by EPSRC (EP/T022213 and EP/W032260).

\bibliography{apssamp}

\end{document}


\title{Characterization of Spin-Orbit Effects in Superconductors In$_5$Bi$_3$ and In$_5$Sb$_3$\\ Supporting Information}
\author{Yao Wei}
\affiliation{Department of Materials Science and Metallurgy, University of Cambridge, 27 Charles Babbage Road, Cambridge CB3 0FS, United Kingdom}
\affiliation{Theory and Simulation of Condensed Matter
(TSCM), King's College London, Strand, London WC2R 2LS, United Kingdom}
\author{Siyu Chen}
\affiliation{Department of Materials Science and Metallurgy, University of Cambridge, 27 Charles Babbage Road, Cambridge CB3 0FS, United Kingdom}

\affiliation{TCM Group, Cavendish Laboratory, University of Cambridge,
J. J. Thomson Avenue, Cambridge CB3 0HE, United Kingdom}
\author{Bartomeu Monserrat}
\email{bm418@cam.ac.uk}
\affiliation{Department of Materials Science and Metallurgy, University of Cambridge, 27 Charles Babbage Road, Cambridge CB3 0FS, United Kingdom}
\affiliation{TCM Group, Cavendish Laboratory, University of Cambridge,
J. J. Thomson Avenue, Cambridge CB3 0HE, United Kingdom}

\maketitle

\section{Pseudopotential selection}

\begin{figure}[h]
\includegraphics[width=0.6\columnwidth]{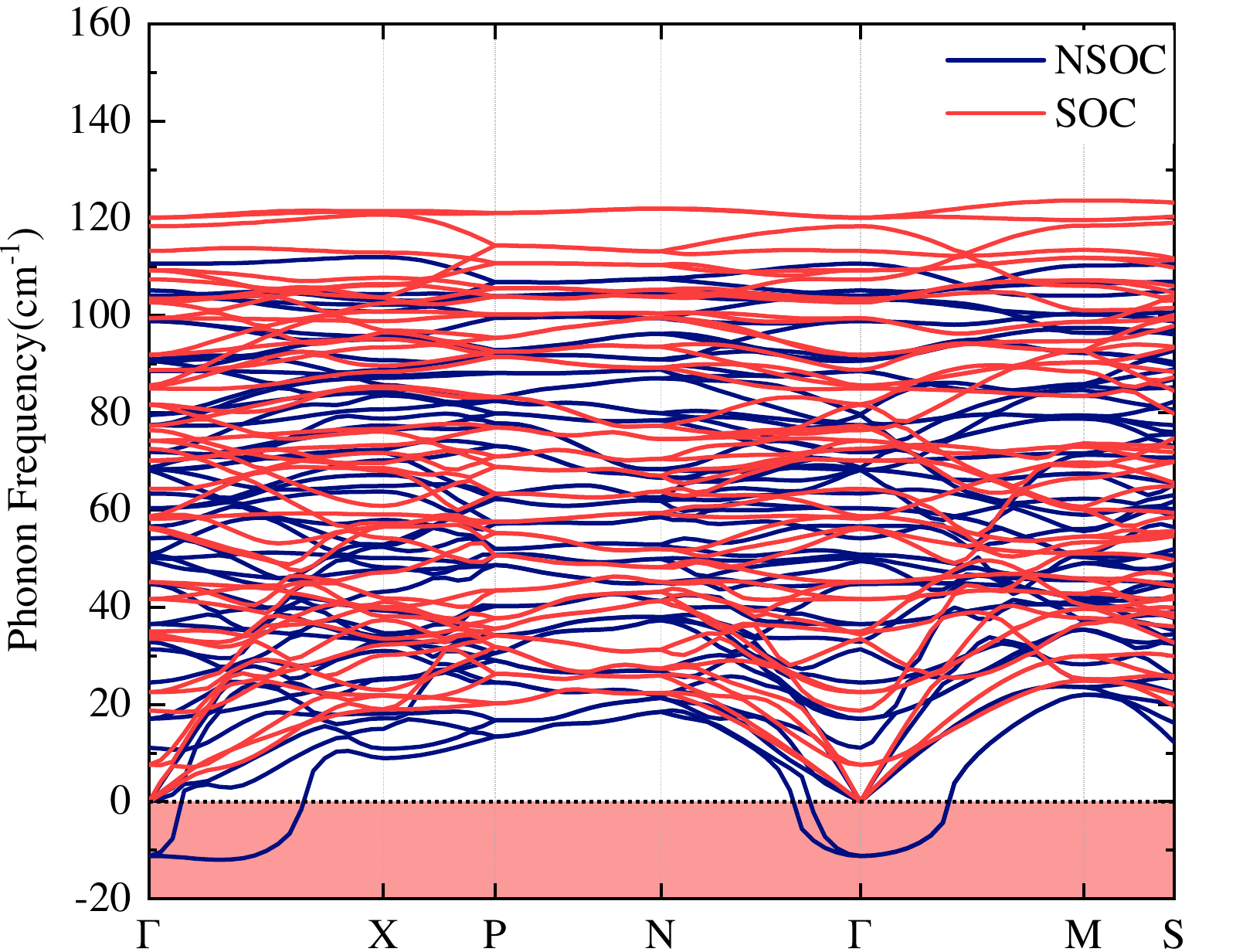}
\caption{Phonon dispersion of  In$_5$Bi$_3$. The results without SOC in dark blue lines and results with SOC in red lines to facilitate comparison, the pseudopotential is pbe-rrkjus-psl}
\label{phonon_dispersion_pp}
\end{figure}

To explore the influence of the pseudopotential on the phonon dispersion
within the {\sc Quantum Espresso} framework, we compare the calculations reported in the main text with those obtained using an alternative pseudopotential
(pbe-dn-rrkjus-psl), and the results are depicted in Fig.\,\ref{phonon_dispersion_pp}. The observed trend closely resembles that described in the main text: without the SOC there are imaginary
phonon frequencies which means that the crystal is dynamically unstable; including
the SOC leads to real phonon frequencies and hence dynamical stability.

\section{Wannier model}

\begin{figure}[ht]
\includegraphics[width=0.6\columnwidth]{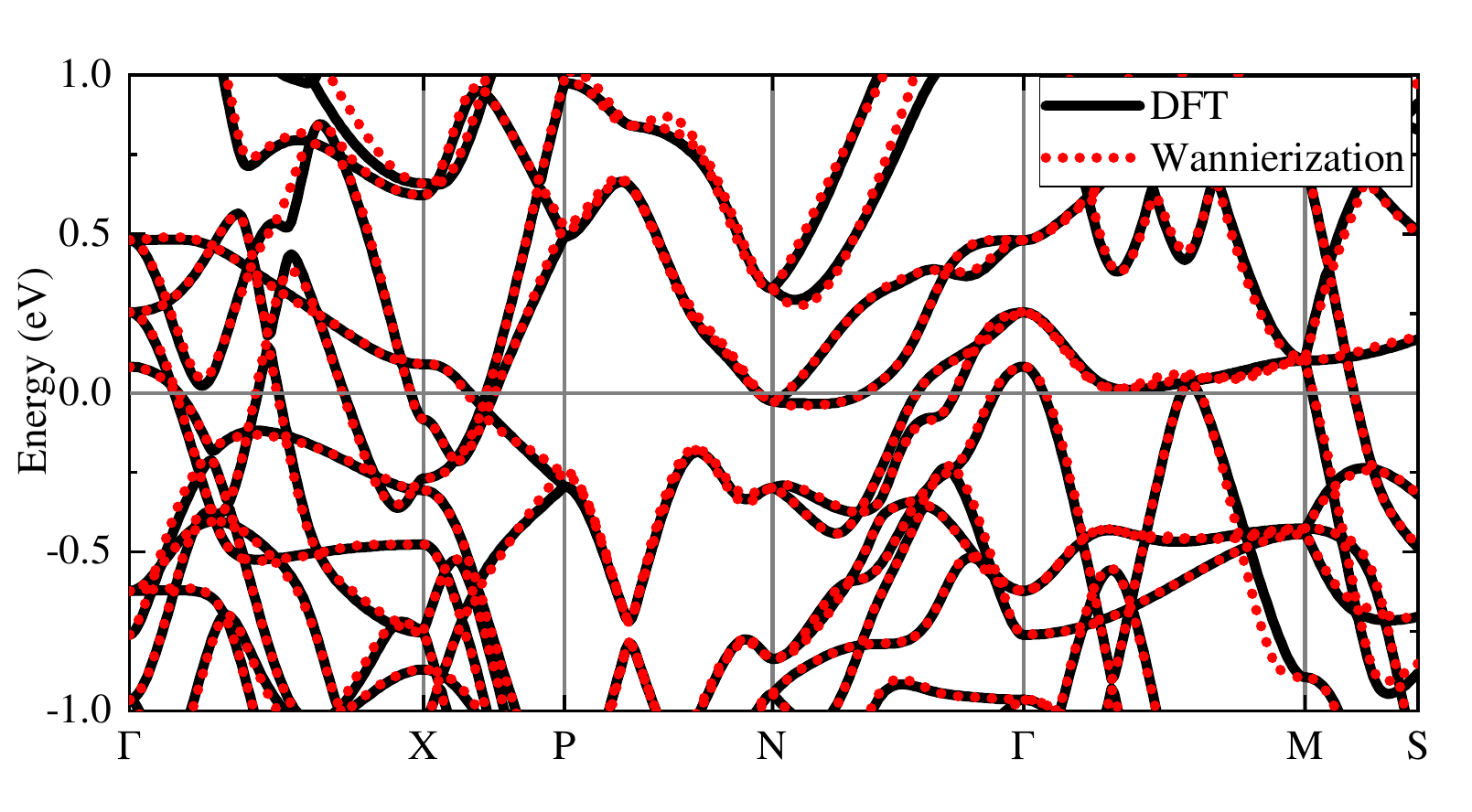}
\caption{The comparison of the electronic structure of In$_{5}$Sb$_{3}$ without
SOC. The Fermi level is at 0.0 eV.}
\label{In5Sb3_nsoc_wannier}
\end{figure}

\begin{figure}[ht]
\includegraphics[width=0.6\columnwidth]{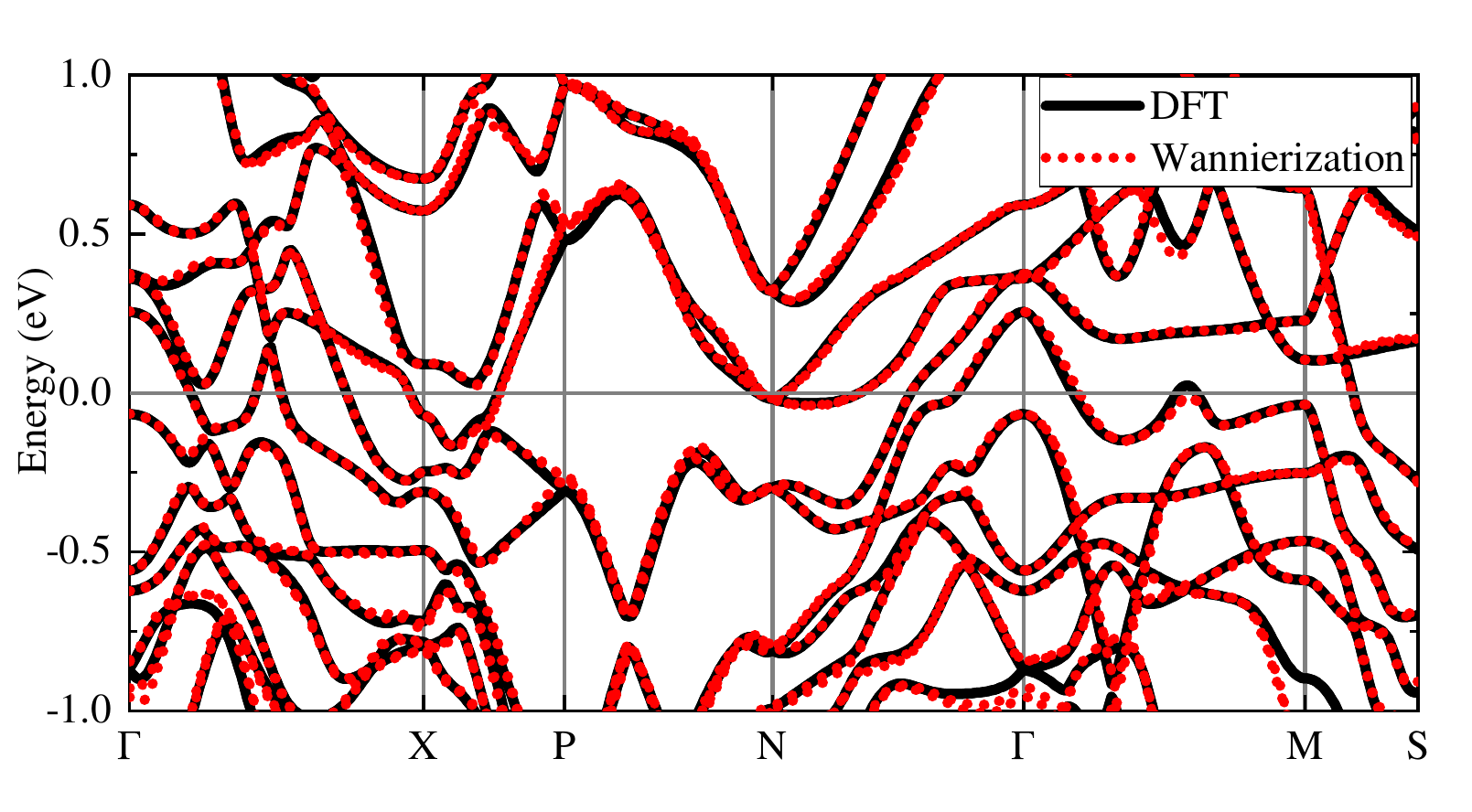}
\caption{The comparison of the electronic structure of In$_{5}$Sb$_{3}$ with
SOC. The Fermi level is at 0.0 eV.}
\label{In5Sb3_soc_wannier}
\end{figure}

\begin{figure}[ht]
\includegraphics[width=0.6\columnwidth]{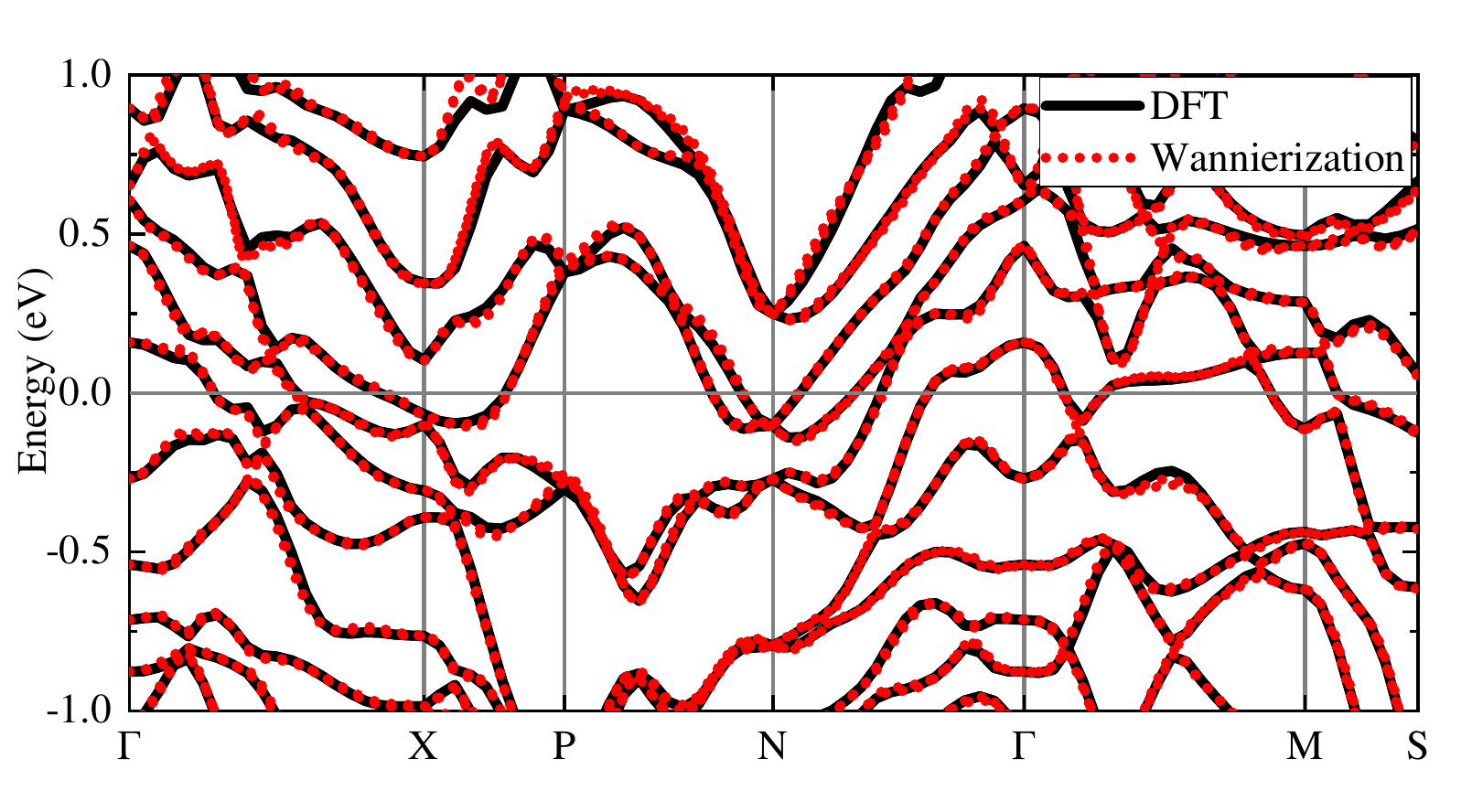}
\caption{The comparison of the electronic structure of In$_{5}$Bi$_{3}$ with
SOC. The Fermi level is set to 0.0 eV.}
\label{In5Bi3_soc_wannier}
\end{figure}

We use the \textsc{wannier90} package\,\citep{mostofi2008wannier90,mostofi2014updated}
to construct Wannier orbitals from the Bloch states (Kohn-Sham
orbitals) of the first principles calculations.
 We give an initial guess in the form of projections on localised orbitals onto Sb \textit{p} and Bi \textit{p} orbitals, and with 2 random orbitals in NSOC calculations, and 14 random orbitals in the SOC calculations.
This procedure enables us to generate a tight-binding Hamiltonian
with a basis of maximally localized Wannier functions\,\cite{marzari1997maximally,souza2001maximally}, which leads to a band structure that closely resembles that obtained from DFT. A well-chosen Wannier window is essential for the accurate calculation of superconducting properties. The key parameters of Wannierization used in this study are presented in Tab.\,\ref{table_wannier}.

\begin{table*}[ht]
\centering
\begin{ruledtabular}
\begin{tabular}{cccc}
Parameter & In$_{5}$Sb$_{3}$ NSOC & In$_{5}$Sb$_{3}$ SOC & In$_{5}$Bi$_{3}$ SOC \\
\hline
nbndsub       & 20   & 50   & 50   \\ 
num\_iter     & 3500 & 3500 & 3500 \\ 
dis\_froz\_max & 9.85 & 10.24 & 11.60 \\ 
dis\_froz\_min & 7.92 & 8.30  & 9.37  \\ 
bands\_skipped & exclude\_bands = 1:65  & exclude\_bands = 1:112  & exclude\_bands = 1:160 \\ 
proj(1)       & random & random & random \\ 
proj(2)       & Sb:p   & Sb:p   & Bi:p   \\ 
\end{tabular}
\end{ruledtabular}
\caption{The key parameters of Wannierization used in this study.}
\label{table_wannier}
\end{table*}

We compare the DFT band structure and the one obtained using Wannier orbitals and the tight-binding both with and without spin-orbit coupling. The results for In$_{5}$Sb$_{3}$ are shown in Figs.\,\ref{In5Sb3_nsoc_wannier} and \ref{In5Sb3_soc_wannier}, while for In$_{5}$Bi$_{3}$ with SOC
the results are shown in Fig.\,\ref{In5Bi3_soc_wannier}.

\section{Convergence of the superconducting transition temperature}

\begin{table}[ht]
\centering
\begin{ruledtabular}
\begin{tabular}{cccc}
$\mathbf{q}$-point grids & $2\times2\times2$ & $3\times3\times3$ & $4\times4\times4$ \\
\hline
McMillan equation  & 1.90 K & 4.02 K & 4.46 K \\
Eliashberg equations & 2.63 K & 5.59 K & 5.63 K \\
\end{tabular}
\end{ruledtabular}
\caption{Predicted superconducting transition temperatures with different $\mathbf{q}$-point grids for In$_{5}$Sb$_{3}$ without SOC, assuming $\mu^{*}$=0.13.}
\label{q2:q_grids_compare}
\end{table}

\begin{table}[ht]
\centering
\begin{ruledtabular}
\begin{tabular}{cccc}
$\mathbf{q}$-point grids & $2\times2\times2$ & $3\times3\times3$ & $4\times4\times4$ \\
\hline
McMillan equation  & 0.76 K & 2.85 K & 3.27 K \\
Eliashberg equations & 1.30 K & 3.53 K & 4.14 K \\
\end{tabular}
\end{ruledtabular}
\caption{Predicted superconducting transition temperatures for In$_{5}$Bi$_{3}$ with SOC, calculated using different $\mathbf{q}$-point grids, assuming $\mu^{*}$=0.13..}
\label{q_grids_compare}
\end{table}

We compare the calculated superconducting transition temperatures for In$_{5}$Sb$_{3}$ without SOC, using $\mathbf{q}$-point samplings of $2\times2\times2$, $3\times3\times3$, and $4\times4\times4$, as shown in Tab.\,\ref{q2:q_grids_compare}. For the $4\times4\times4$ $\mathbf{q}$-point grid, the transition temperature is 5.63 K, determined using the Eliashberg equation. When using the $3\times3\times3$ grid, the temperature is only 0.04 K lower, indicating that the $3\times3\times3$ $\mathbf{q}$-point sampling is sufficiently accurate for predicting the transition temperature in NSOC calculations. However, using the $2\times2\times2$ grid results in a $T_{c}$ that is 2 K lower.

We also performed a comparison for In$_{5}$Bi$_{3}$ with SOC, using $\mathbf{q}$-point grids of $2\times2\times2$, $3\times3\times3$, and $4\times4\times4$, as shown in Tab.\,\ref{q_grids_compare}. For the $4\times4\times4$ grid, the transition temperature is approximately 0.6 K higher than for the $3\times3\times3$ grid, indicating that the electron-phonon coupling calculation has not yet converged at $3\times3\times3$ $\mathbf{q}$-point sampling in SOC calculations. This highlights the need for higher precision in such calculations.

\bibliography{apssamp}